\documentclass[12pt]{book}

\usepackage[dvips]{graphicx,color}
\usepackage{makeidx,hori,zon}
\makeauthorindex

\newcommand{\lsim}   {\mathrel{\mathop{\kern 0pt \rlap
  {\raise.2ex\hbox{$<$}}}
  \lower.9ex\hbox{\kern-.190em $\sim$}}}
\newcommand{\gsim}   {\mathrel{\mathop{\kern 0pt \rlap
  {\raise.2ex\hbox{$>$}}}
  \lower.9ex\hbox{\kern-.190em $\sim$}}}
\def\beq{\begin{equation}}
\def\eeq{\end{equation}}
\def\ba{\begin{eqnarray}}
\def\ea{\end{eqnarray}}

\BookTitle{Inflating Horizon of Particle Astrophysics and Cosmology}
\CopyRight{\copyright 2005 by Universal Academy Press, Inc.\ and 
Yamada Science Foundation}

\begin{document}

\pagenumbering{arabic}

\chapter{Dip in UHECR and Transition from Galactic to Extragalactic Cosmic Rays}

\author{Veniamin Berezinsky\\
{\it INFN, Laboratori Nazionali del Gran Sasso, 67010 Assergi (AQ), Italy}}
\AuthorContents{V.\ Berezinsky}
\AuthorIndex{Berezinsky}{V.}

\section*{Abstract}
The dip is a feature in the diffuse spectrum of UHE protons
in energy range $1\times 10^{18} - 4\times 10^{19}$~ eV, which 
is caused by electron-positron pair production on CMB photons. 
Calculated for power-law generation spectrum with index $\gamma_g=2.7$,
the shape of the dip is confirmed with high accuracy by data of 
Akeno - AGASA, HiRes, Yakutsk and Fly's Eye detectors. The predicted
shape of the dip is robust: it is valid for the rectilinear and
diffusive propagation, for different discretenesses in the source 
distribution, for local source overdensity and deficit, for source 
inhomogeneities on scale $\ell \lsim 100$~Mpc etc. Below the
characteristic energy $E_c \approx 1\times 10^{18}$~eV the spectrum of 
the dip flattens for both diffusive and rectilinear propagation, and
more steep galactic spectrum becomes dominant 
at $E < E_c$. The energy of transition $E_{\rm tr} < E_c$ 
approximately coincides with the position of the second knee $E_{2kn}$ 
observed in the cosmic ray spectrum. The critical energy $E_c$ is
determined by the energy $E_{\rm eq} = 2.3\times 10^{18}$~eV, 
where adiabatic and pair-production energy losses are equal. Thus,
position of the second knee is explained by proton energy losses on
CMB photons. 

\section{Introduction}
\label{introduction}
The nature of signal carriers of UHECR is not yet
established. The most natural primary particles are extragalactic
protons. Due to interaction with the CMB radiation the UHE protons
from extragalactic sources are predicted to have a sharp
steepening of energy spectrum, so called GZK cutoff \cite{GZK}.

There are two other signatures of extragalactic protons in the
spectrum: dip and bump \cite{HS85} -  \cite{Stanev00}.
The dip is produced due to
$p+\gamma_{\rm CMB} \to p+e^++e^-$ interaction.
The bump is produced by pile-up protons
which loose energy in the GZK cutoff. As was demonstrated in
\cite{BG88}, the
bump is clearly seen in the calculations for a single source at 
large redshift $z$, but
it practically disappears in the diffuse spectrum, because individual
peaks are located at different energies.

As it will be discussed in this paper, the dip is a reliable feature in the
UHE proton spectrum (see also \cite{BGG} - \cite{BGG3}).
Being relatively faint feature, it is however clearly seen in the
spectra observed by AGASA, Fly's Eye,
HiRes and Yakutsk arrays (see \cite{agasa} - \cite{NW} for the data).
We argue here that it can be
considered as the confirmed signature of interaction of extragalactic
UHE protons with CMB.

The dip has low-energy and high-energy flattenings (see Fig~\ref{mfactor}). 
The high-energy
flattening at $E_a \approx 1\times 10^{19}$~eV  reproduces the {\em ankle},  
the feature well seen in the observational data. The low-energy
flattening at  $E_c \approx 1\times 10^{18}$~eV, seen for both 
rectilinear \cite{BGG} and diffusive \cite{AB,Lem,AB1} propagation,
is a natural place for transition from extragalactic to galactic
component of cosmic rays. Somewhere below 
$E_c$ the more steep galactic
component must appear. Thus $E_c$ describes the beginning of transition 
(if one follows it from high energy side) or the energy where this
transition completes (if one moves from low to high energy). 
We shall demonstrate here (see also \cite{AB1}) that $E_c$ is
connected with $E_{\rm eq} = 2.3 \times 10^{18}$~eV, where 
pair-production and adiabatic  energy losses are equal.  The visible
transition from galactic to extragalactic cosmic rays occurs at 
$E_{\rm tr} < E_c$, and this energy coincides well with position of
the {\em second knee}  (Akeno - $6\times 10^{17}$~eV, Fly's Eye - 
$4\times 10^{17}$~eV, HiRes - $7\times 10^{17}$~eV and 
Yakutsk - $8\times 10^{17}$~eV). 

The transition from galactic to extragalactic cosmic rays at 
$E_c \approx 1\times 10^{18}$ implies the dominance of the proton 
composition of the 
observed cosmic rays at $E \gsim 1\times 10^{18}$~eV. While 
HiRes \cite{mass-Hires} HiRes-MIA \cite{Hi-Mia} and Yakutsk \cite{Glushkov00} 
data support this prediction and
Haverah-Park \cite{HP} data do not contradict it at 
$E \gsim (1 - 2)\times 10^{18}$~eV,
the Akeno \cite{mass-Akeno} and  Fly's Eye \cite{FE} data favour the mixed 
composition, dominated by heavy nuclei (for a review see \cite{NW} and 
\cite{Watson04}). 

Below we shall analyze the features in UHE proton spectrum using
basically two assumptions: the uniform distribution of the sources in the
universe and the power-law generation spectrum. We do not consider the 
possible speculations, such as cosmological evolution of sources. 
Only in Section \ref{AGN} we shall turn to model-dependent consideration. 

\section{The dip}
\label{thedip}
The analysis of the dip is convenient to perform in terms of the
{\em modification factor}.\\
Modification factor is defined as a ratio of the spectrum $J_p(E)$ 
with all energy losses taken into account, and unmodified
spectrum $J_p^{\rm unm}$, where only adiabatic energy losses (red shift) are
included.
\beq  
\eta(E)=\frac{J_p(E)}{J_p^{\rm unm}(E)}.  
\label{modif}  
\eeq  
 \begin{figure}[ht]
  \begin{center}
    \includegraphics[width=8.0cm]{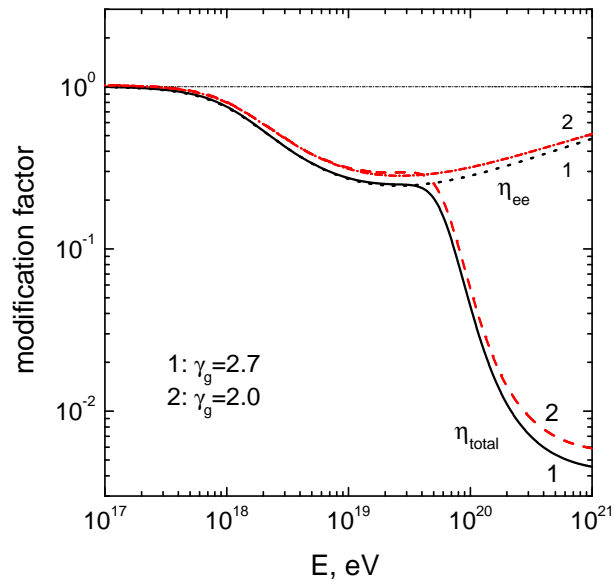}
\end{center}
  \caption{Modification factor for the power-law generation spectra
with $\gamma_g$ in the range 2.0 - 2.7. Curve $\eta=1$ corresponds to
adiabatic energy losses, curves $\eta_{ee}$ - to adiabatic and pair
production energy losses and curves $\eta_{tot}$ - to total energy losses.} 
  \label{mfactor}
\end{figure}
Modification factor is less model-dependent quantity than the
spectrum. In particular, it depends weakly on $\gamma_g$, because
both numerator and denominator in Eq.~(\ref{modif}) include
$E^{-\gamma_g}$.
In this paper we shall consider the non-evolutionary case
$m=0$.
In Fig~\ref{mfactor} the modification factors are shown as a function
of energy for two spectrum indices $\gamma_g=2.0$ and $\gamma_g=2.7$.
They do not differ much. 

\begin{figure}[ht]
\begin{minipage}[h]{8cm}
\centering
\includegraphics[width=7.6cm,clip]{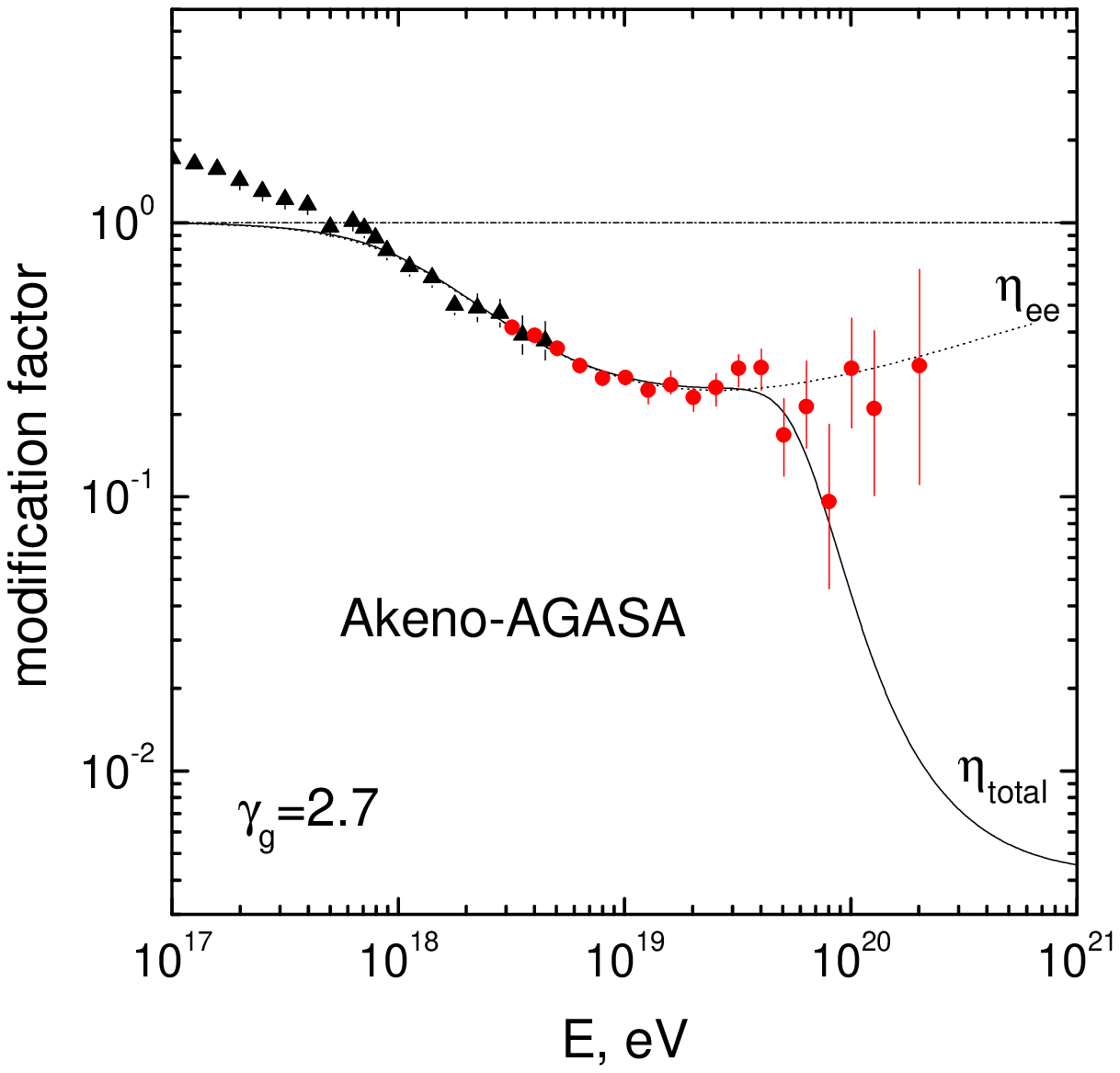}
\end{minipage}
\hspace{2mm}
\begin{minipage}[h]{8cm}
\centering
\includegraphics[width=7.6cm,clip]{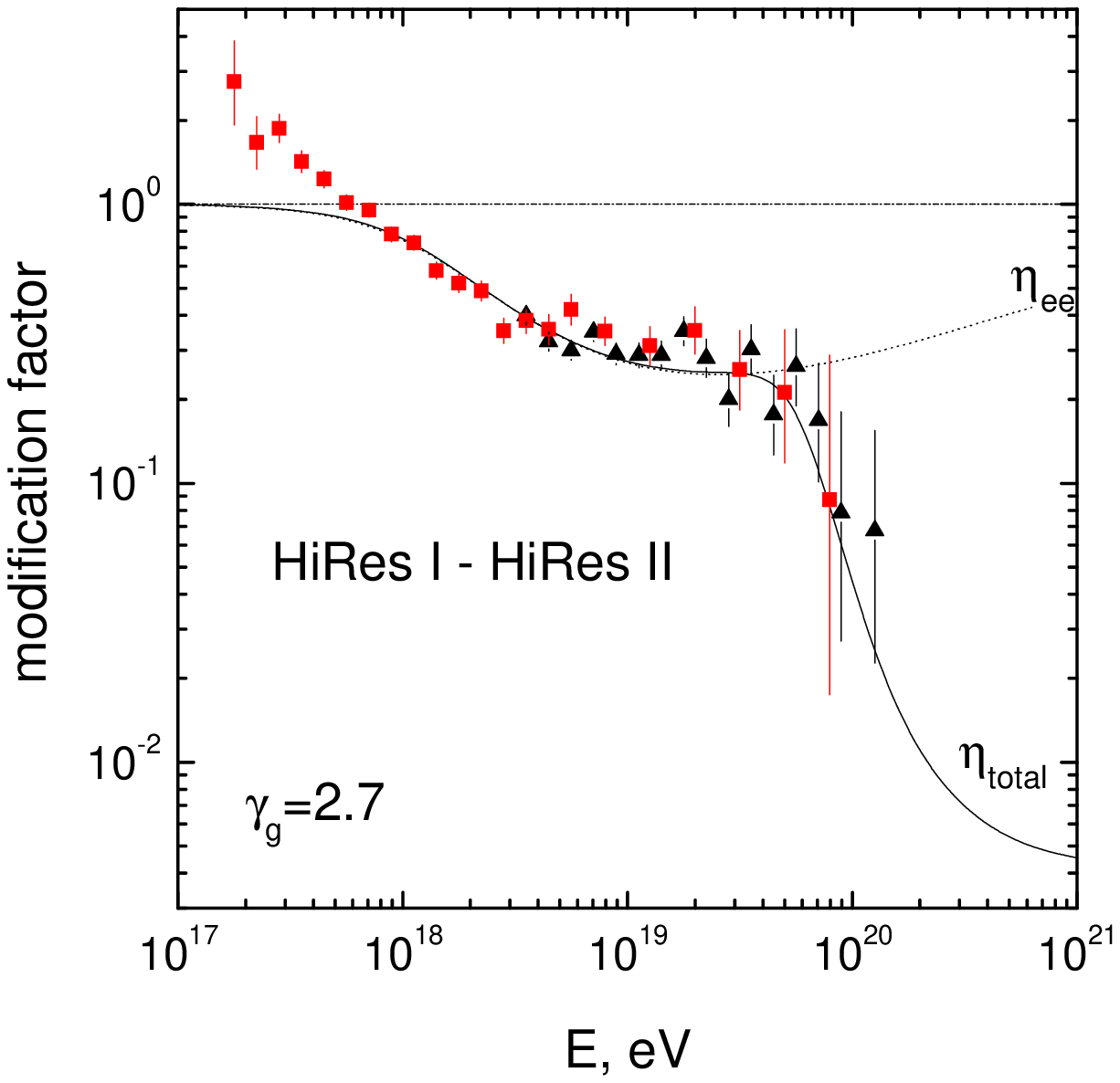}
\end{minipage}
\vspace{2mm}
\begin{minipage}{8cm}
\centering
\includegraphics[width=7.6cm,clip]{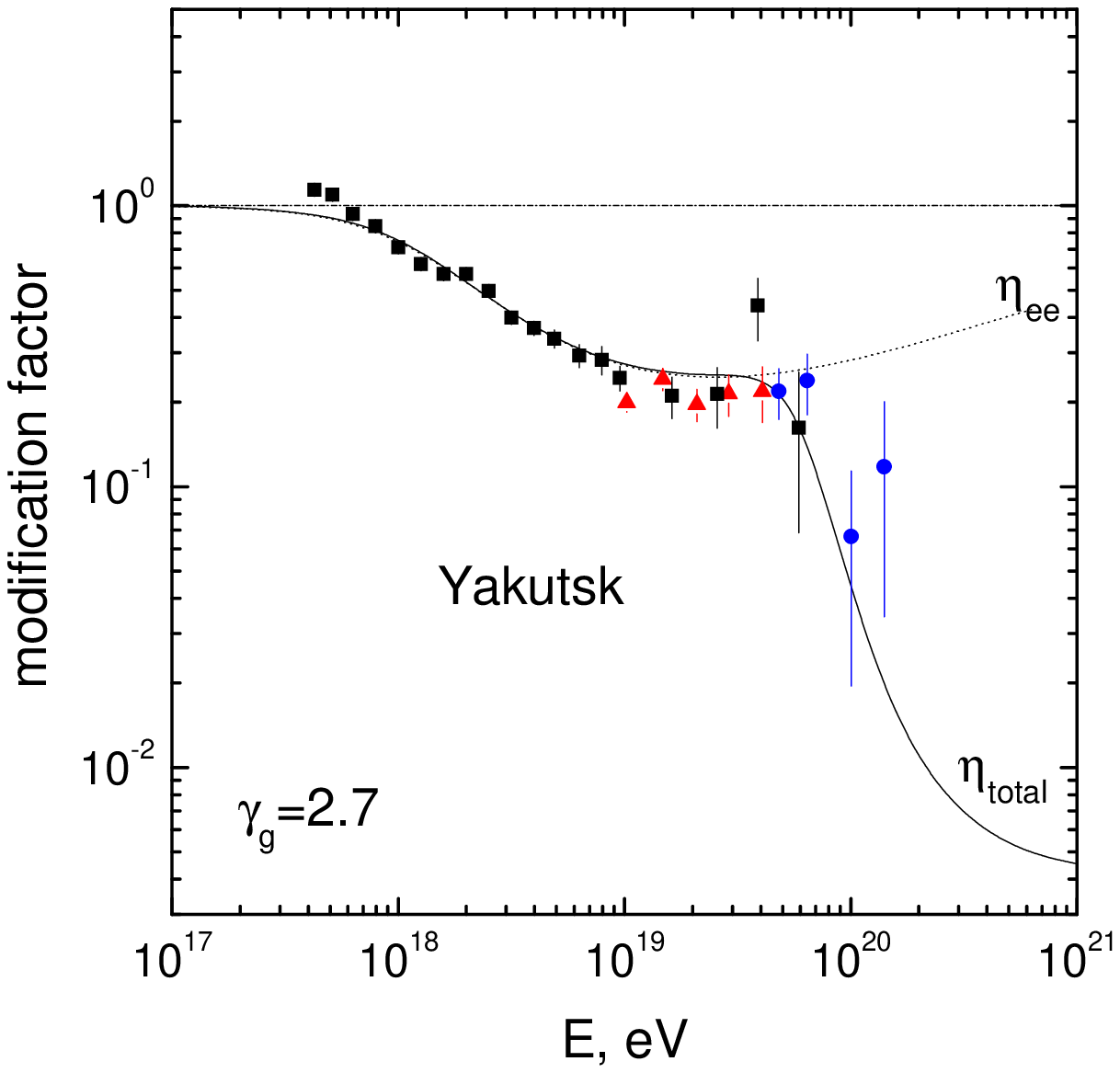}
\end{minipage}
\hspace{19mm} 
\begin{minipage}[h]{8cm}
\centering
\includegraphics[width=7.4cm,height=7.4cm,clip]{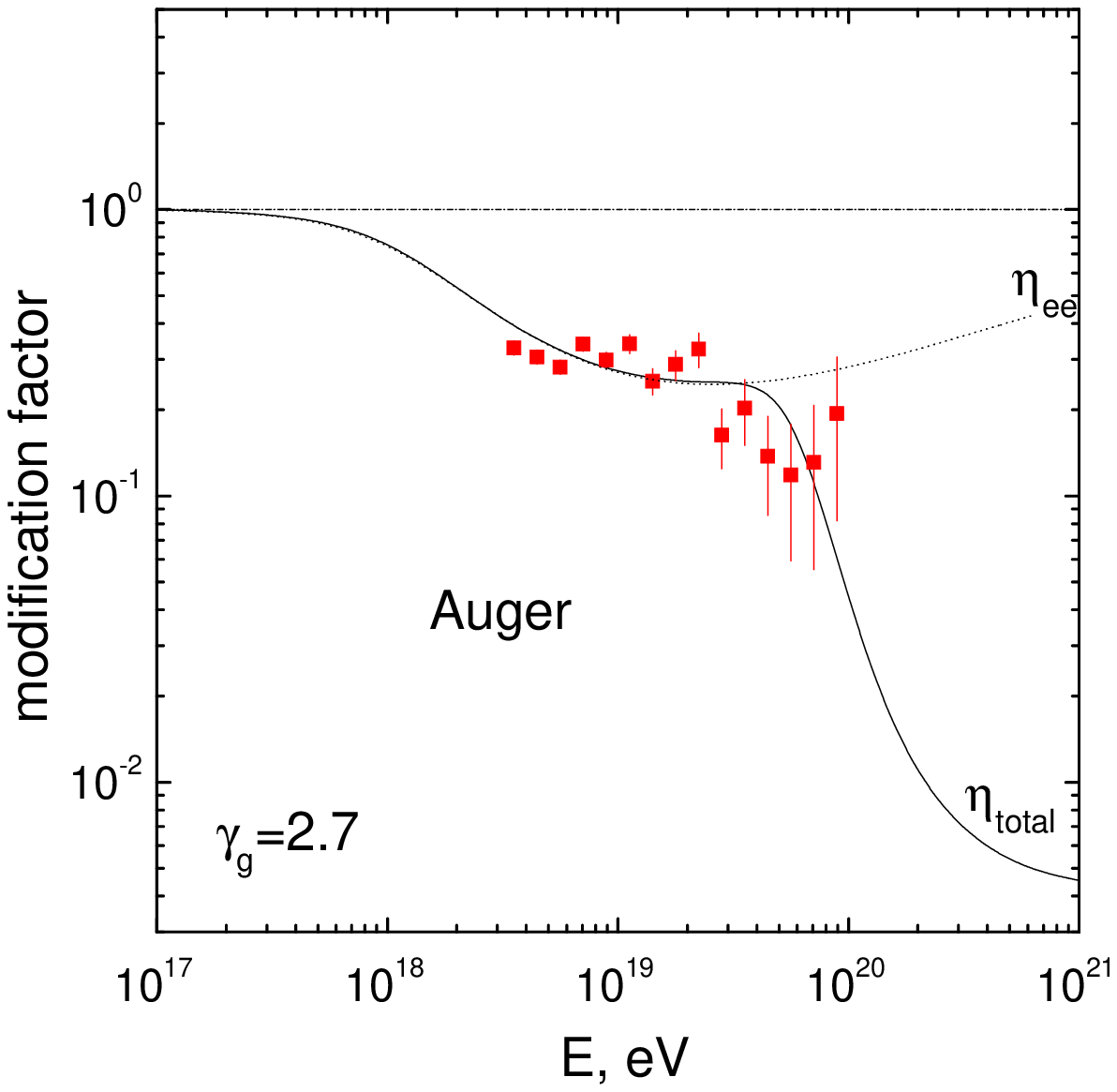}
\end{minipage}
\caption{\label{dips} Predicted dip in comparison with AGASA, HiRes, 
Yakutsk and Auger\protect\cite{Auger} data.}
\end{figure}

Comparison of the predicted dip ($\eta_{ee}$ curve)  with observational 
data are shown in Fig~\ref{dips} for $\gamma_g=2.7$.
One can observe the excellent agreement with AGASA, HiRes and Yakutsk
data (and also with Fly's Eye data not presented here), while the
Auger spectrum, at this preliminary stage, does not contradict the dip.

\begin{figure}[ht]
\begin{minipage}[h]{8cm}
\centering
\includegraphics[width=71mm,height=71mm,clip]{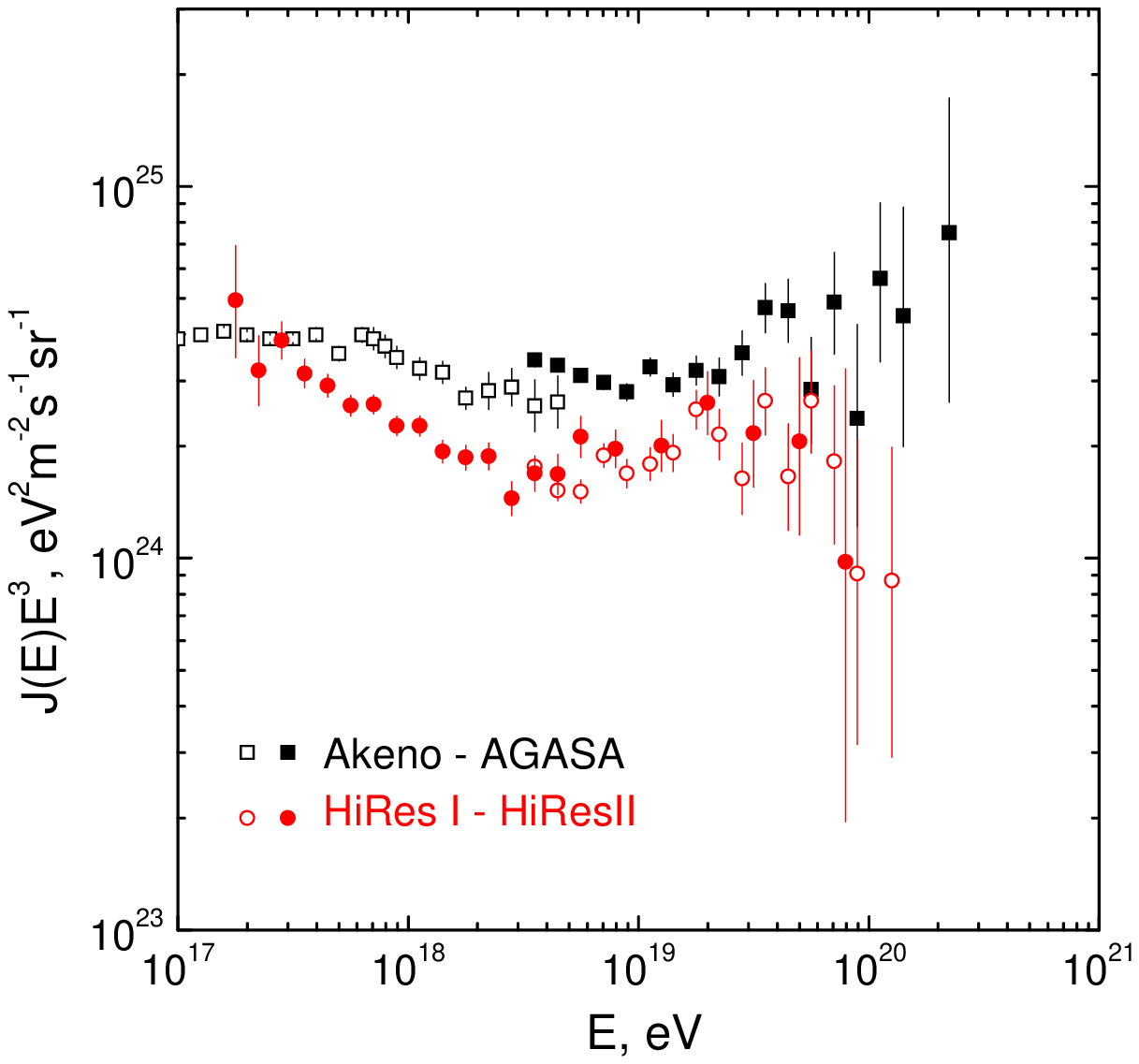}
\end{minipage}
\hspace{5mm}
\begin{minipage}[h]{8cm}
\centering
\includegraphics[width=7.6cm,clip]{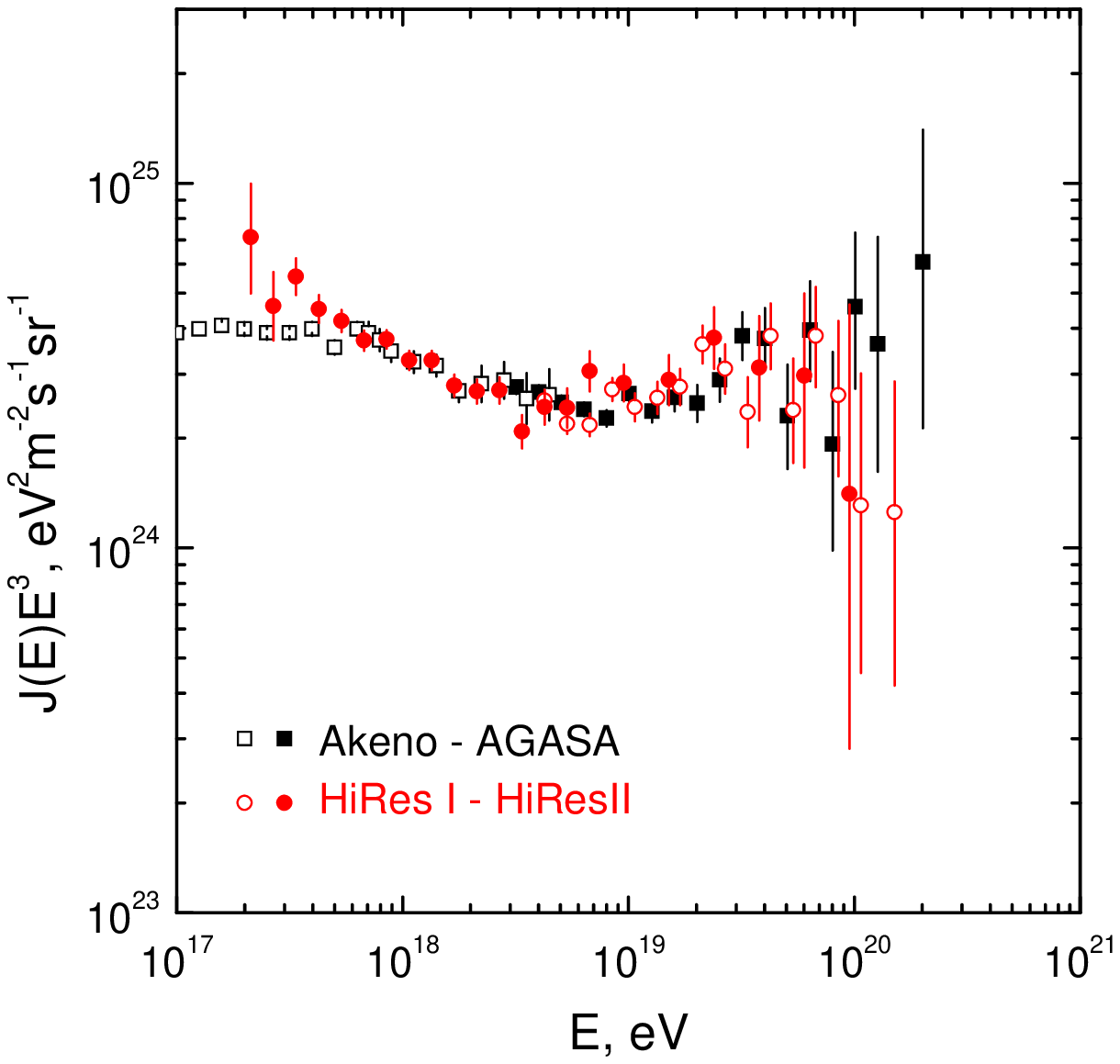}
\end{minipage}
\caption{\label{Ag-Hi} Spectra of Akeno-AGASA and HiRes (left
 panel) and  with energy
  calibrated by dip (right panel). 
}
\end{figure}

The systematic errors in energy determination of existing detectors 
exceed 20\%. The dip can be used for energy calibration of
detectors. Assuming the energy-independent systematic errors we 
shifted the energies of AGASA and HiRes by factor $\lambda_{\rm Ag}=0.9$
and  factor $\lambda_{\rm Hi}=1.2$, respectively, to reach the best fit
of the dip.  In Fig~\ref{Ag-Hi} the spectra of
Akeno-AGASA and HiRes are shown before and after this
energy calibration. One can see the good agreement of the data 
at $E < 1\times 10^{20}$~eV and their consistency at 
$E > 1\times 10^{20}$~eV. This result should be considered together 
with calculations \cite{DBO}, where it was demonstrated that 
11 superGZK AGASA events can be simulated by the spectrum with GZK
cutoff in case of 30\% error in energy determination. We may
tentatively conclude that existing discrepancy between AGASA, HiRes
and Auger spectra at all energies may have statistical origin. 

\noindent
{\em 2.1 Robustness and caveats}\\*[2mm]
How robust is prediction of the dip and what are uncertainties? 

\begin{figure}[ht]
\begin{minipage}[h]{8cm}
\centering
\includegraphics[width=7.6cm,clip]{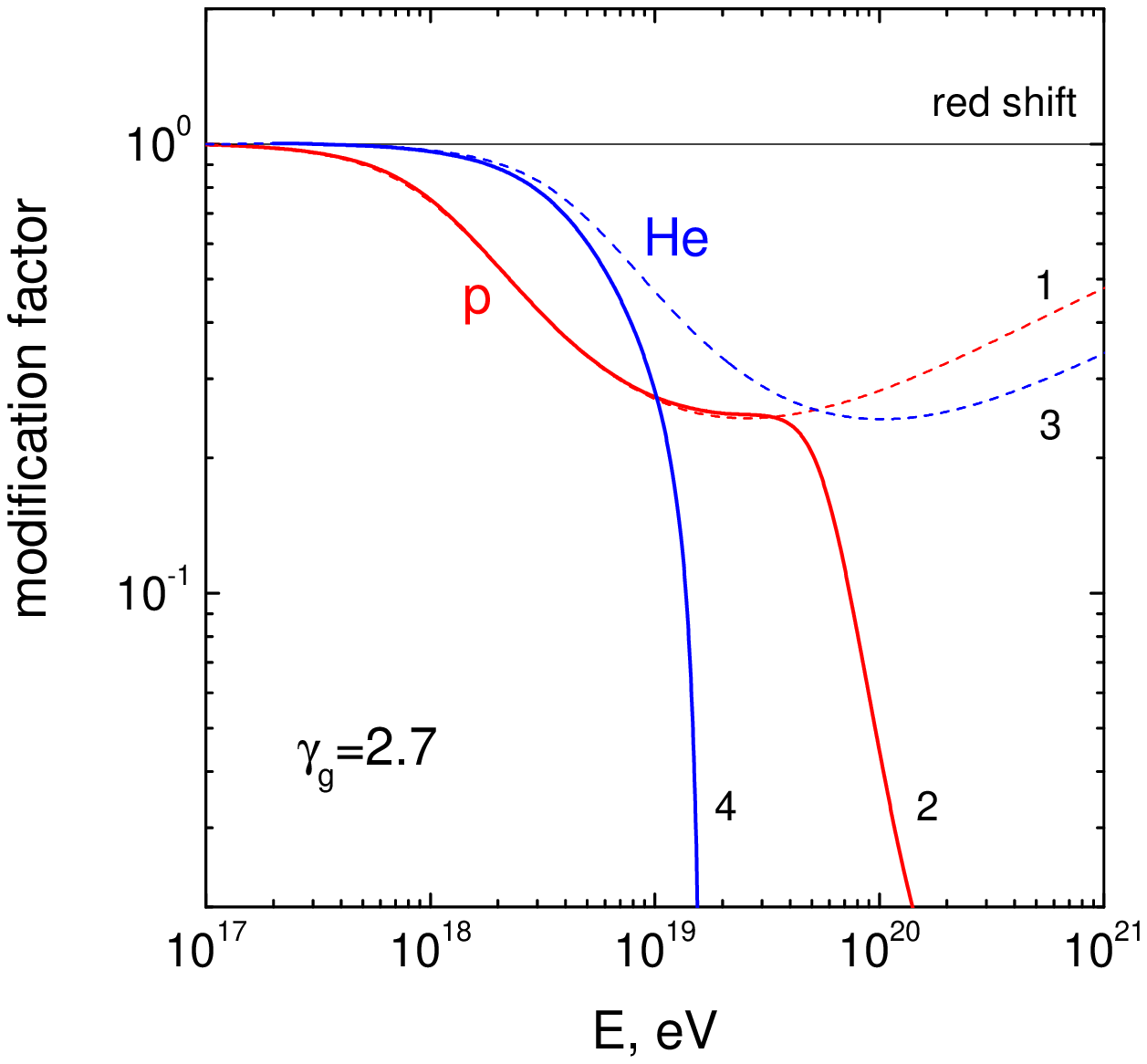}
\end{minipage}
\hspace{5mm}
\begin{minipage}[h]{8cm}
\centering
\includegraphics[width=7.6cm,clip]{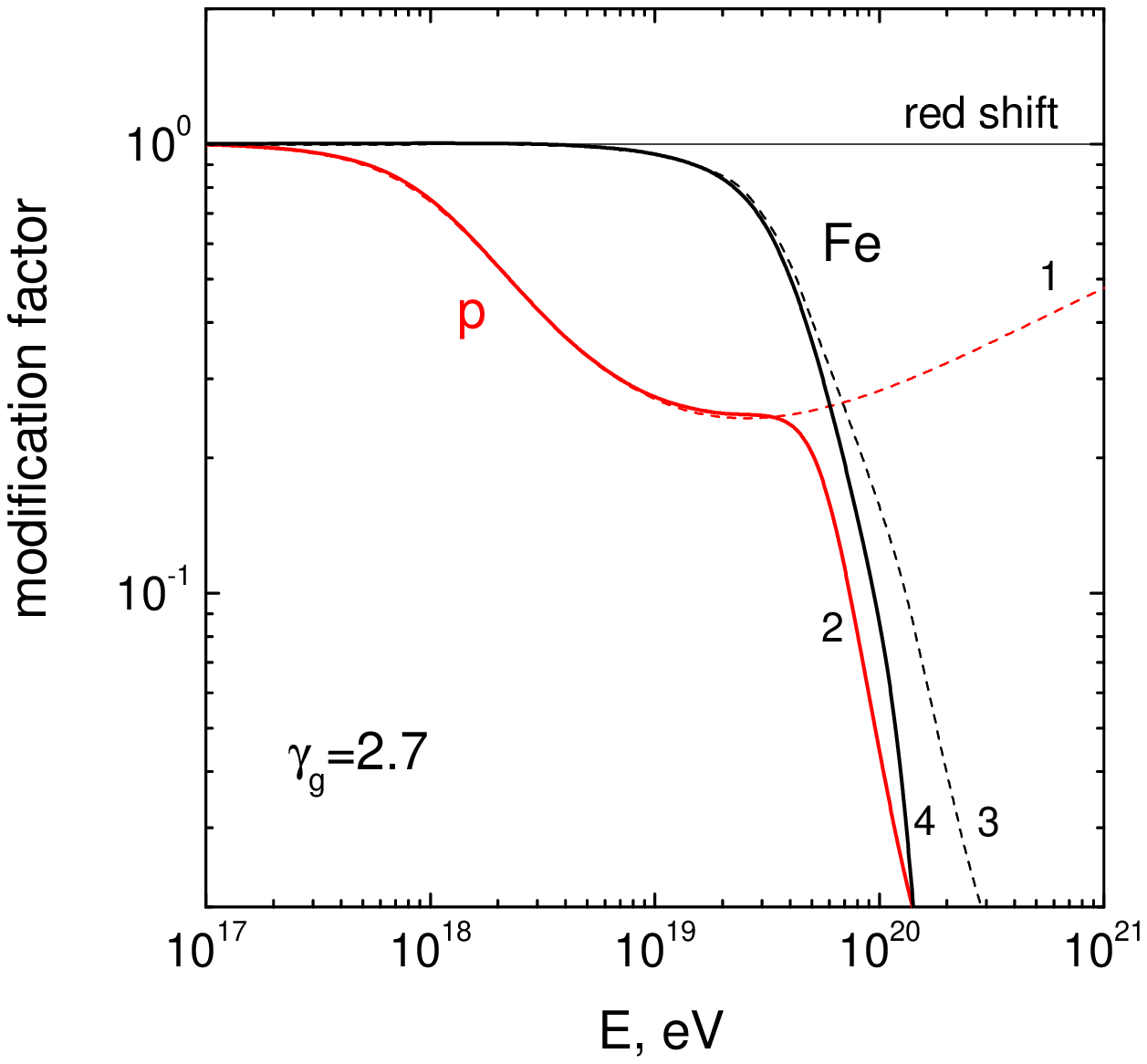}
\end{minipage}
\caption{\label{dip-nucl} Modification factors for helium and iron
 nuclei in comparison with that for protons. Proton modification
 factors are given by curves 1 and 2.
 Nuclei modification factors are
 given by curves 3 (adiabatic and pair production energy losses) and
 by curves 4 (with photodisintegration included).
 }
\end{figure}

The shape of the calculated proton dip is stable relative to many phenomena
included in calculations, namely, discreteness in the source
distribution, mode of propagation (diffusive and rectilinear),
inhomogeneities in source distribution on the scale $\ell \lsim 100$~Mpc,  
local source overdensity and deficit, acceleration $E_{\rm max}$ and 
fluctuations in energy losses. However, it is modified by the assumption
of source evolution \cite{BGG} and by presence of UHE nuclei in primary
flux \cite{BGG3} (see also \cite{Allard}). In Fig~\ref{dip-nucl} the dips 
for helium and iron nuclei are presented in comparison with the proton dip. 
One can see that presence of 10 - 20 \% of 
nuclei in the primary flux upsets the good agreement of proton dip
with observations. Below we shall study this problem in a more
detailed way.  

Consider some acceleration mechanism operating in a gas with 
the mixed space density of hydrogen, $n_H$, and nuclei with atomic  
number A , $n_A$. We assume the power-law generation spectrum 
\beq
Q_A(p)=K_A n_A p^{-\gamma_g}, 
\label{gen}
\eeq
where $\gamma_g=2.7$ and minimum momentum is $p_A^{\rm min}$ for
nuclei A. We assume also that the total number of particles 
accelerated per unit time $Q_A^{\rm tot}=K n_A$, where $K$ is 
independent of $A$ and of charge number $Z$. Then one obtains for the 
ratio of generation rates of $A$ nuclei and protons   
\begin{equation}
Q_A(p)/Q_p(p)=\left\{ \begin{array}{ll}
(n_A/n_H)A^{\gamma_g-1} ~&{\rm if}~~ p_A^{\rm min}=
v_{\rm min}\Gamma_{\rm min} A m_N\\ 
(n_A/n_H)Z^{\gamma_g-1} ~&{\rm if}~~ p_A^{\rm min}=R_{\rm min}Z
\end{array}
\right. ,
\label{ratio}
\end{equation}
where $\Gamma$ is Lorentz factor and $R=p/Z$ is rigidity. We shall
refer to the upper case in Eq.~(\ref{ratio}) as to $\Gamma$-acceleration 
(it includes the shock acceleration) and to the lower case as to rigidity 
acceleration. 

For rigidity acceleration in single-ionized gas, $Z_{\rm eff}=1$, one has
\beq
Q_A(p)/Q_p(p)= n_A/n_H
\label{ratio-ioniz}
\eeq
This ionization condition is important only near threshold of 
acceleration $p\gsim p_A^{\rm min}$, at higher energies the gas can be
fully ionized 

The modification factor for the mixed composition is given by 
\beq
\eta (E)= \frac{\eta_p(E)+\lambda \eta_A(E)}{1+\lambda},
\label{mod-mix}
\eeq
where $\lambda=Q_A^{\rm unm}(p)/Q_p^{\rm unm}(p)$. The strongest
distortion of proton modification factor $\eta_p(E)$ is given by
helium nuclei for which $n_A/n_H=0.06$, corresponding to mass ratio 
$Y_p=0.24$.  
\begin{figure}[ht]
\begin{minipage}[h]{8cm}
\centering
\includegraphics[width=7.6cm,clip]{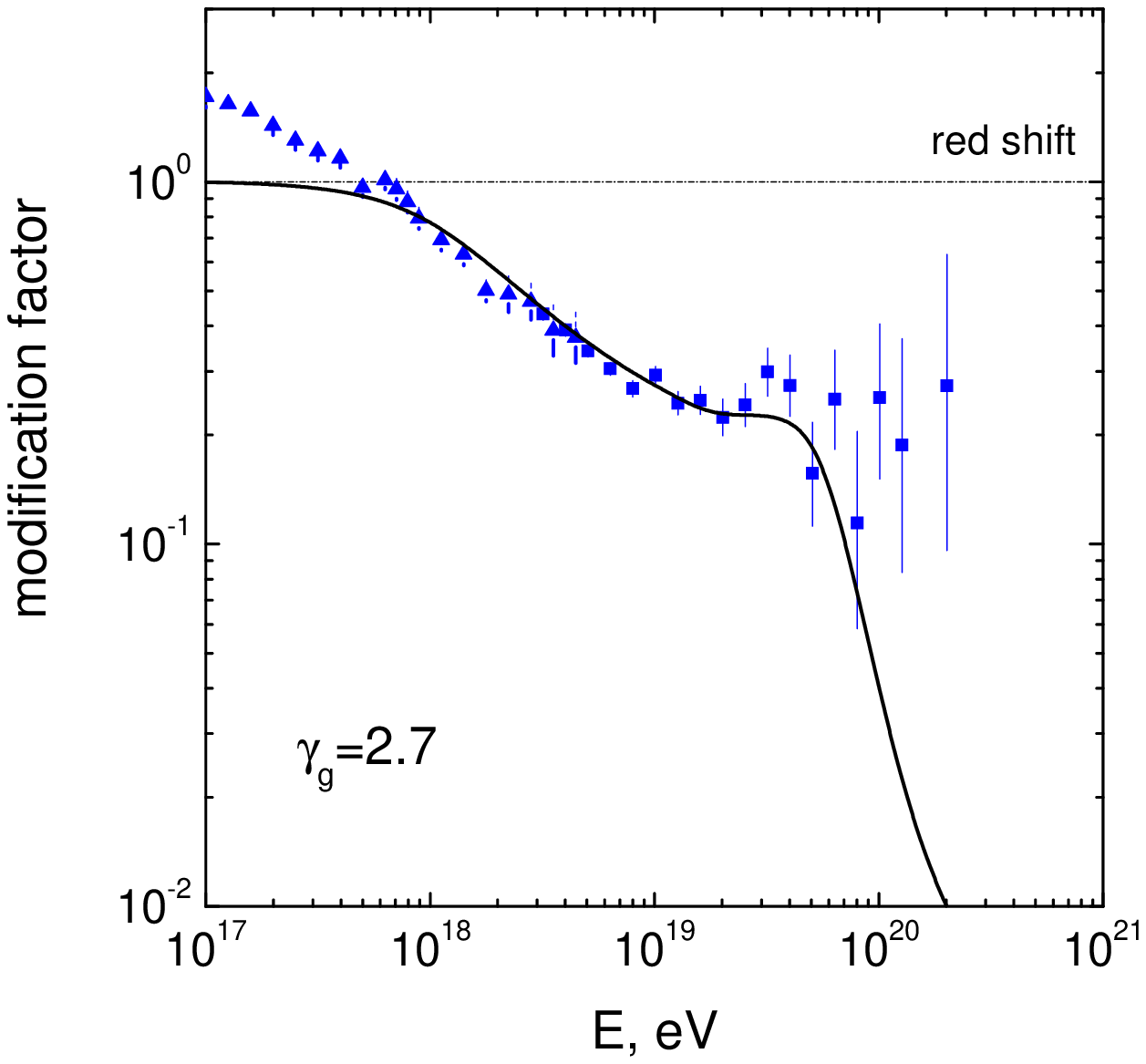}
\end{minipage}
\hspace{5mm}
\begin{minipage}[h]{8cm}
\centering
\includegraphics[width=7.6cm,clip]{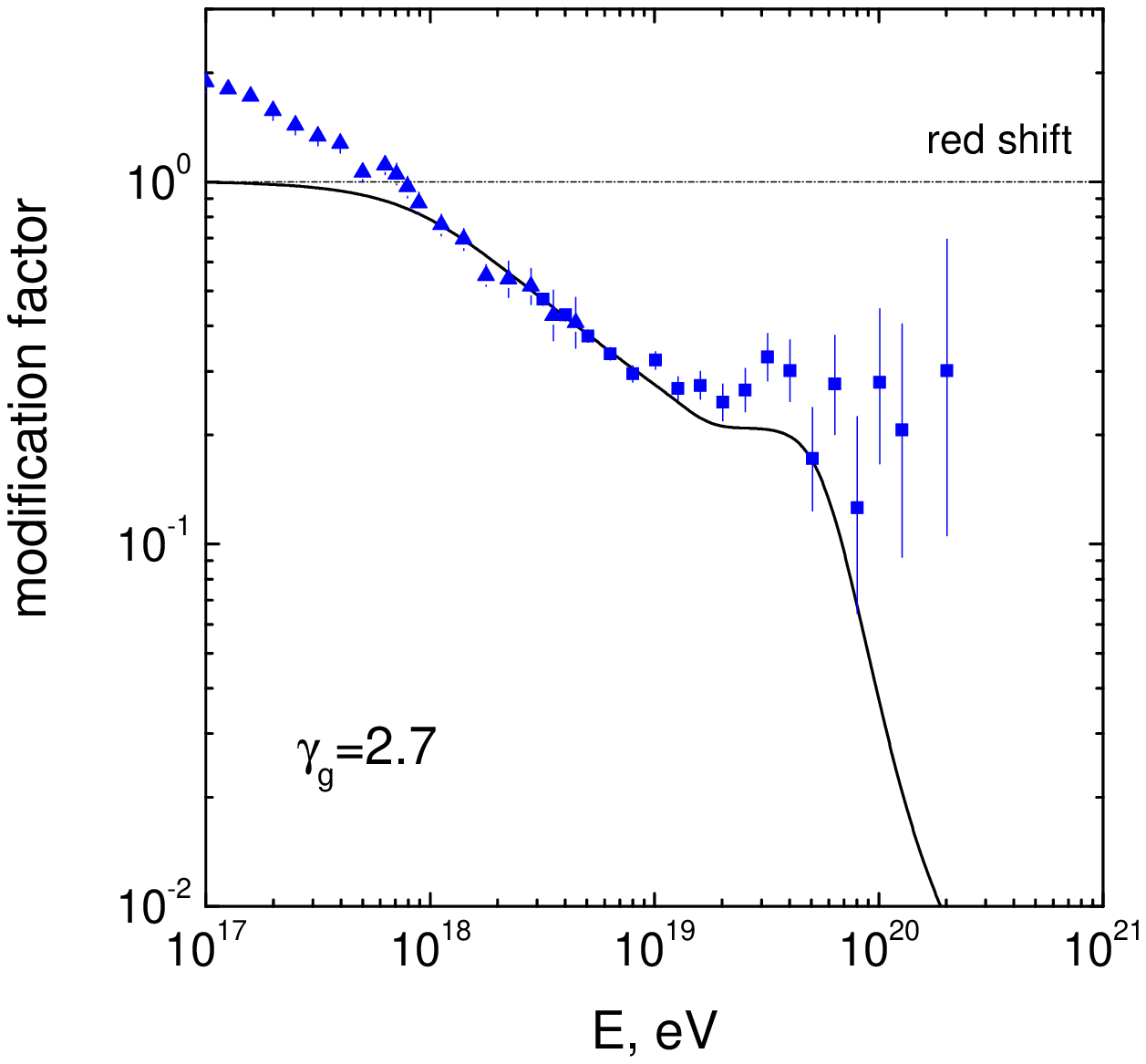}
\end{minipage}
\caption{\label{dip-mix} Modification factors for the mixed composition
of protons and helium nuclei in comparison with AGASA data. The left
panel corresponds to mixing parameter $\lambda=0.1$, and the right
panel to $\lambda=0.195$.
}
\end{figure}
Fig~\ref{dip-mix} shows that rigidity acceleration in single ionized 
helium gas with mixing parameter $\lambda=0.06$ agrees well with the
data, and in fully ionized helium gas with $\lambda=0.195$ 
(lower case in Eq.~\ref{ratio}) agrees 
worse. We shall remind that the first and the second ionization potential
for helium is very high, 24.6 and 54.4~eV, respectively. In extreme case of
neutral helium, the mixed modification factor $\eta(E) \approx \eta_p(E)$. 
$\Gamma$-acceleration, and shock acceleration in particular, results
in bad agreement of the p+He dip with observations. 
One should keep in mind the 
approximate character of our estimates, especially concerning the 
assumption about power-law spectrum down to $p_{\rm min}$.

If agreement of the proton dip with observations is not accidental 
(the probability of this is small according to small $\chi^2$/d.o.f.),
Fig~\ref{dip-mix} should be interpreted as indication to possible 
acceleration mechanism. It cannot be shock acceleration, which results 
in large $\lambda \propto A^{\gamma_g -1}$. The large $\gamma_g$ 
demands large $E_{\rm min}$ in acceleration mechanism to avoid too large 
cosmic ray luminosity of a source. Correlation with BL Lacs indicates 
the jet acceleration (see Section~\ref{AGN}). 

Another option is given by photodisintegration of UHE nuclei by
radiation. The relevant calculations have been recently performed in 
\cite{Sigl}. It was demonstrated that for the magnetized sources the
observed flux above $1\times 10^{19}$~eV becomes pure proton one due to 
photodisintegration by CMB radiation. We can add here that nuclei can be 
destroyed at lower energies by infra-red radiation inside a source. 
An example of such model can be given by acceleration in the inner jet, 
where helium nuclei can be photodisintegrated by infra-red emission 
from accretion disc.   
\section{Transition from extragalactic to galactic cosmic rays}
\label{transition}
\begin{figure}[ht]
\begin{minipage}[h]{8cm}
\centering
\includegraphics[width=7.6cm,clip]{mfactorAGASA.eps}
\end{minipage}
\begin{minipage}[h]{8cm}
\centering
\includegraphics[width=72mm,height=72mm,clip]{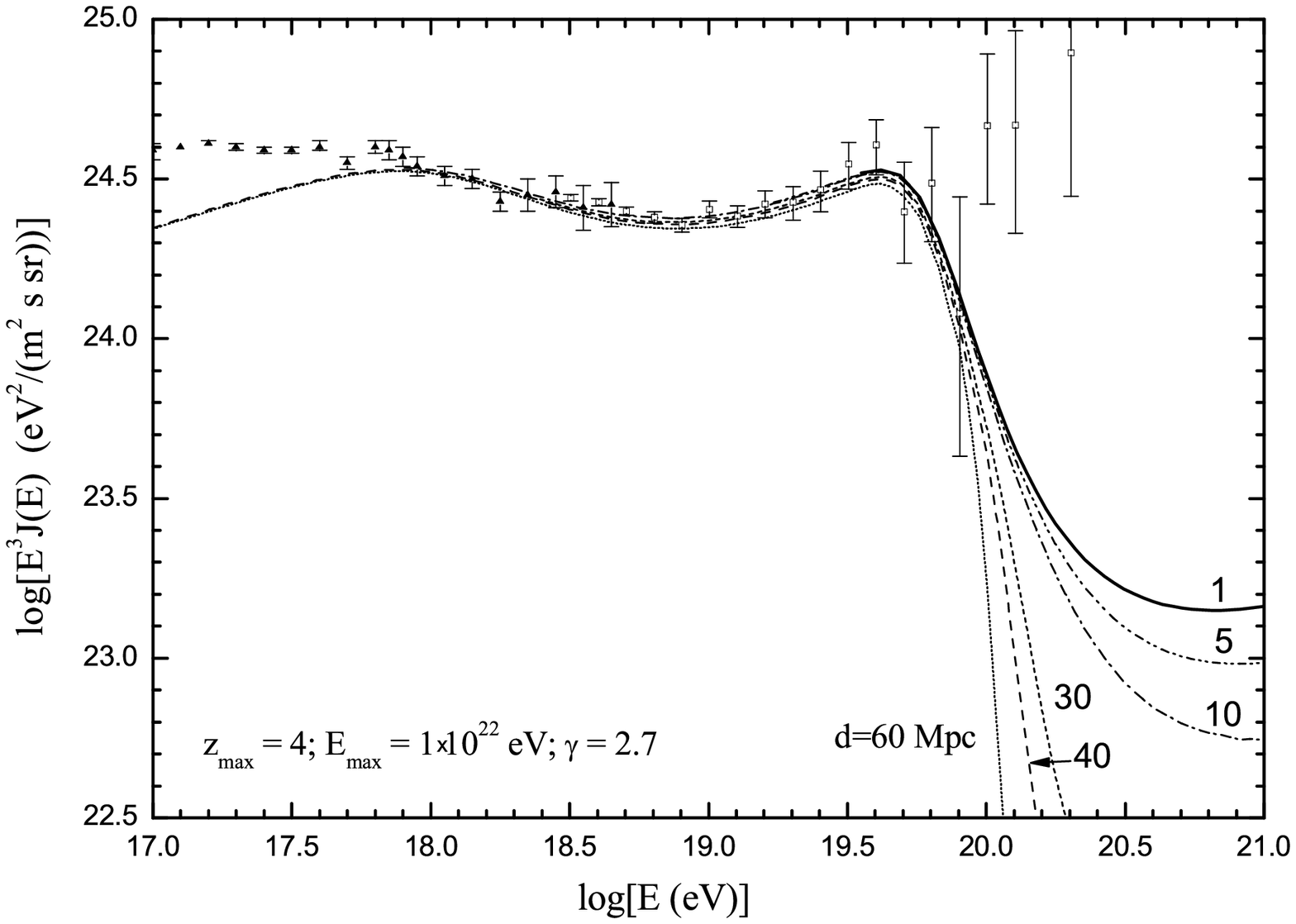}
\end{minipage}
\vspace{2mm}
\begin{minipage}{8cm}
\centering
\includegraphics[width=7.6cm,clip]{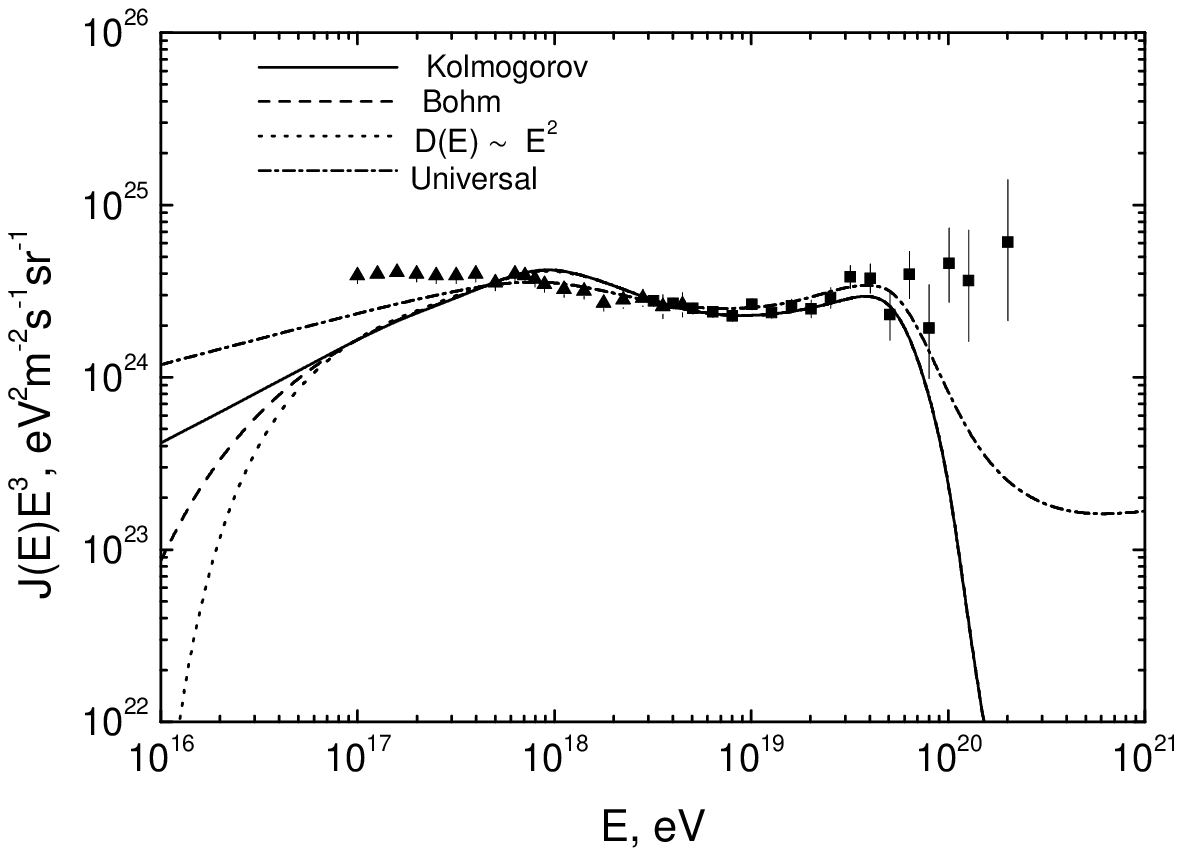}
\end{minipage}
\hspace{15mm} 
\begin{minipage}[h]{8cm}
 \caption{\label{trans} Appearance of transition energy 
$E_c \approx 1\times 10^{18}$~eV in the modification factor compared
  with AGASA data (left panel), in the spectra for
  rectilinear propagation for the sources with separation $d$
  indicated in the figure (right panel) and in the spectra 
  for diffusive propagation (lower panel) 
  }
\end{minipage}
\end{figure}

We will follow this transition going from high to low energies. 
The critical energy $E_c \approx 1\times 10^{18}$~eV  is given by the 
low-energy steepening of the dip (see Fig~\ref{mfactor}), below which
the more steep galactic spectrum becomes dominant. The data presented
in Fig~\ref{trans} confirm this expectation.  

In the left panel the
modification factor is shown in comparison with AGASA data (see also 
Fig~\ref{dips} for HiRes). At $E < E_c$  
the experimental modification factor exceeds 1, while by definition 
(see Eq.~\ref{modif}) $\eta (E) \leq 1$. It signals about appearance of a
new component at $E < 1\times 10^{18}$~eV. 

In the right panel the spectrum for rectilinear propagation from the
sources with different separation $d$ and with $\gamma_g=2.7$ is compared with 
AGASA data. One can see that at $E < 1\times 10^{18}$~eV the
calculated extragalactic spectrum becomes less than that observed. 
In the lower panel the similar comparison is shown for diffusive
propagation with different diffusion regimes at lower energies. The
dash-dotted curve (universal spectrum) corresponds to the case when
the separation between sources $d \to 0$. In all cases the
transition again occurs at $E_c \approx 1\times 10^{18}$~eV. 

What is the reason of this universality? 
\subsection{$E_{\rm eq}$,~~ $E_c$~~ and the second knee $E_{\rm 2kn}$}
\label{E_c}
We study the transition, moving from high towards low energies. 
$E_c$ is the beginning of transition (or its end, if one moves from
low energies). $E_c$ is determined by energy 
$E_{\rm eq}= 2.3\times 10^{18}$~eV, where adiabatic and  pair-production energy
losses become equal. The quantitative analysis of this connection is
given in \cite{AB1}. We shall give here the semi-quantitative explanation. 

The flattening of the spectrum occurs at energies $E \leq E_c$, where 
$E_c=E_{\rm eq}/(1+z_{\rm eff})^2$ and $z_{\rm eff}$ should be
estimated as  redshift
up to which the main contribution to unmodified spectrum occurs.  
The simplified analytic estimate for $\gamma_g=2.6 - 2.8$ gives 
$1+z_{\rm eff} = 1.5$ and hence $E_c \approx 1\times 10^{18}$~eV. In
fact, the right and lower panels of Fig~\ref{trans} presents the
exact result of calculations of this kind. 

In experimental data the transition is searched for as a feature
started at some low energy $E_{\rm 2kn}$ - the second knee. Its
determination depends on experimental procedure, and all what
we can predict is $E_{\rm 2kn} < E_c$. Determined in different experiments
$E_{\rm 2kn} \sim (0.4 - 0.8)\times 10^{18}$~eV. 

The transition at the second knee appears also from consideration of 
propagation of cosmic rays in the Galaxy (see e.g.\cite{2knee}).

Being thought of as purely galactic feature, the position of the
second knee in our analysis appears as direct consequence of
extragalactic proton energy energy losses. 

\subsection{KASCADE data and transition at the ankle}
\label{ankle}
Ankle at energy $E_a \approx 1\times 10^{19}$~eV is seen  
as flattening of the spectrum in data of AGASA, HiRes and Yakutsk
detectors in Fig~\ref{dips}  In many works \cite{ankle} it is considered as 
a position of transition from galactic to extragalactic cosmic rays
(see also \cite{DeMSt} for general analysis of the transition). 
The KASCADE data \cite{kascade} give a clue for understanding of transition.
They confirm the rigidity model, according to which position of a knee
for nuclei with charge Z is connected with the position of the proton
knee $E_p$ as $E_Z=Z E_p$. There are two versions of this model. One
is the confinement-rigidity model (bending above the knee is due to 
insufficient confinement in galactic magnetic field), and the other is 
acceleration-rigidity model ($E_{\rm max}$ is determined by rigidity). 
In both models the heaviest nuclei (iron) 
start to disappear at $E > E_{\rm Fe}= 6.5\times 10^{16}$~eV. 
{How the gap between $1\times 10^{17}$~eV and  $1\times 10^{19}$~eV 
is filled?}

Protons start to disappear at $E > 2.5\times 10^{15}$~eV. {Where they 
came from at $E> 1\times 10^{17}$~eV to be seen in the Akeno detector?}

To ameliorate these problems, the authors of \cite{ankle} shift the ankle to 
$3\times 10^{18}$~eV or assume another galactic component, which
appears after the iron knee. The second-knee model suggests a solution
as transition (being completed) at $E_c \approx 1\times 10^{18}$~eV.  

The ankle model needs acceleration by galactic sources up to 
$1\times 10^{19}$~eV (at least for iron nuclei). The second knee model
ameliorates this requirement by one order of magnitude. 

The second-knee model predicts the spectrum shape down to 
$1\times 10^{18}$~eV with extremely good accuracy ($\chi^2$/d.o.f.= 1.12 
for Akeno-AGASA and $\chi^2$/d.o.f.= 0.908 for HiRes). In the ankle
model one has to consider this agreement as accidental, though such 
hypothesis has very small probability, determined by $\chi^2$ cited
above. 
\begin{figure}[ht]
\begin{minipage}[h]{9cm}
\centering
\includegraphics[width=86mm,height=70mm,clip]{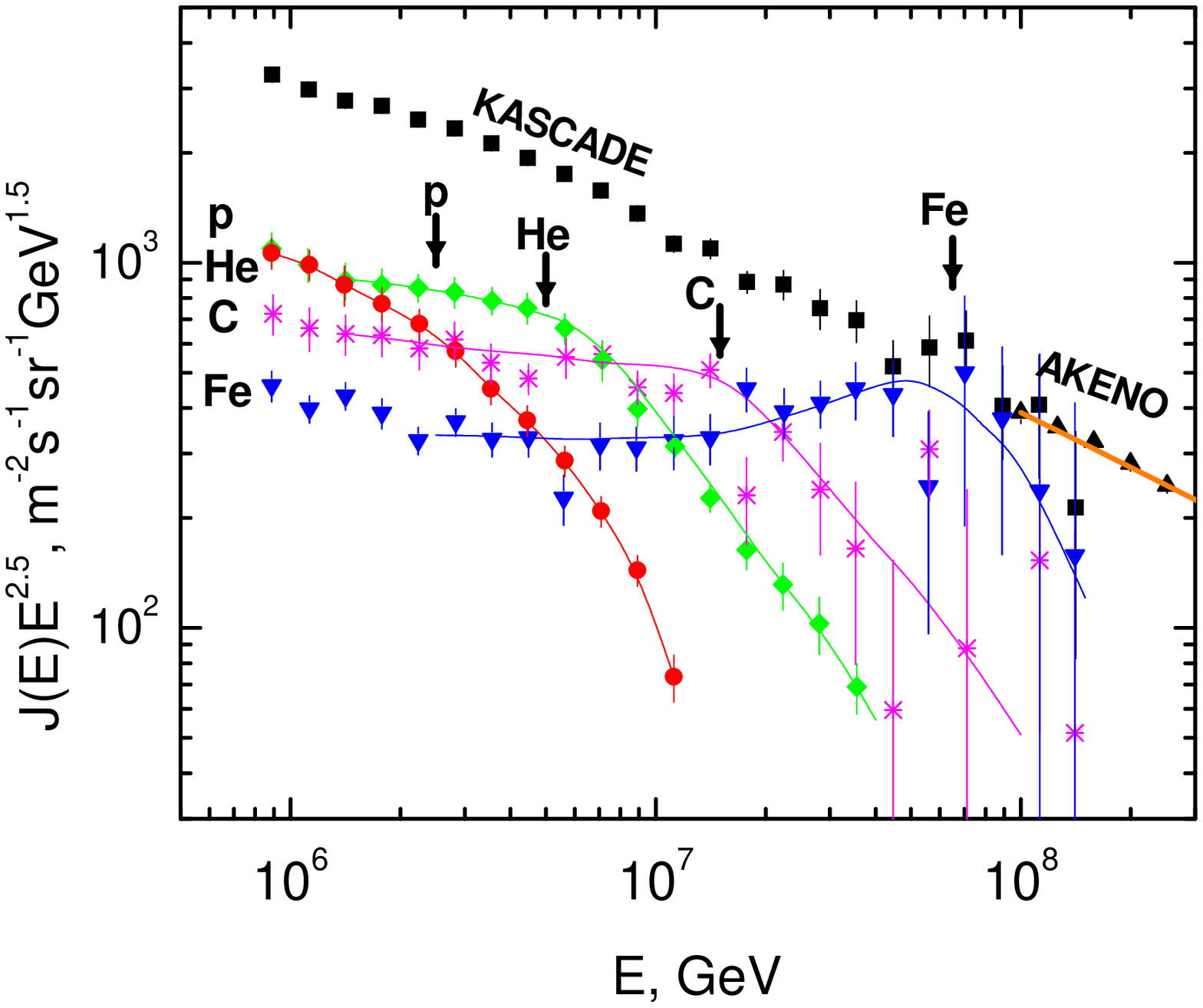}
\end{minipage}
\begin{minipage}[h]{9cm}
\centering
\includegraphics[width=86mm,height=70mm,clip]{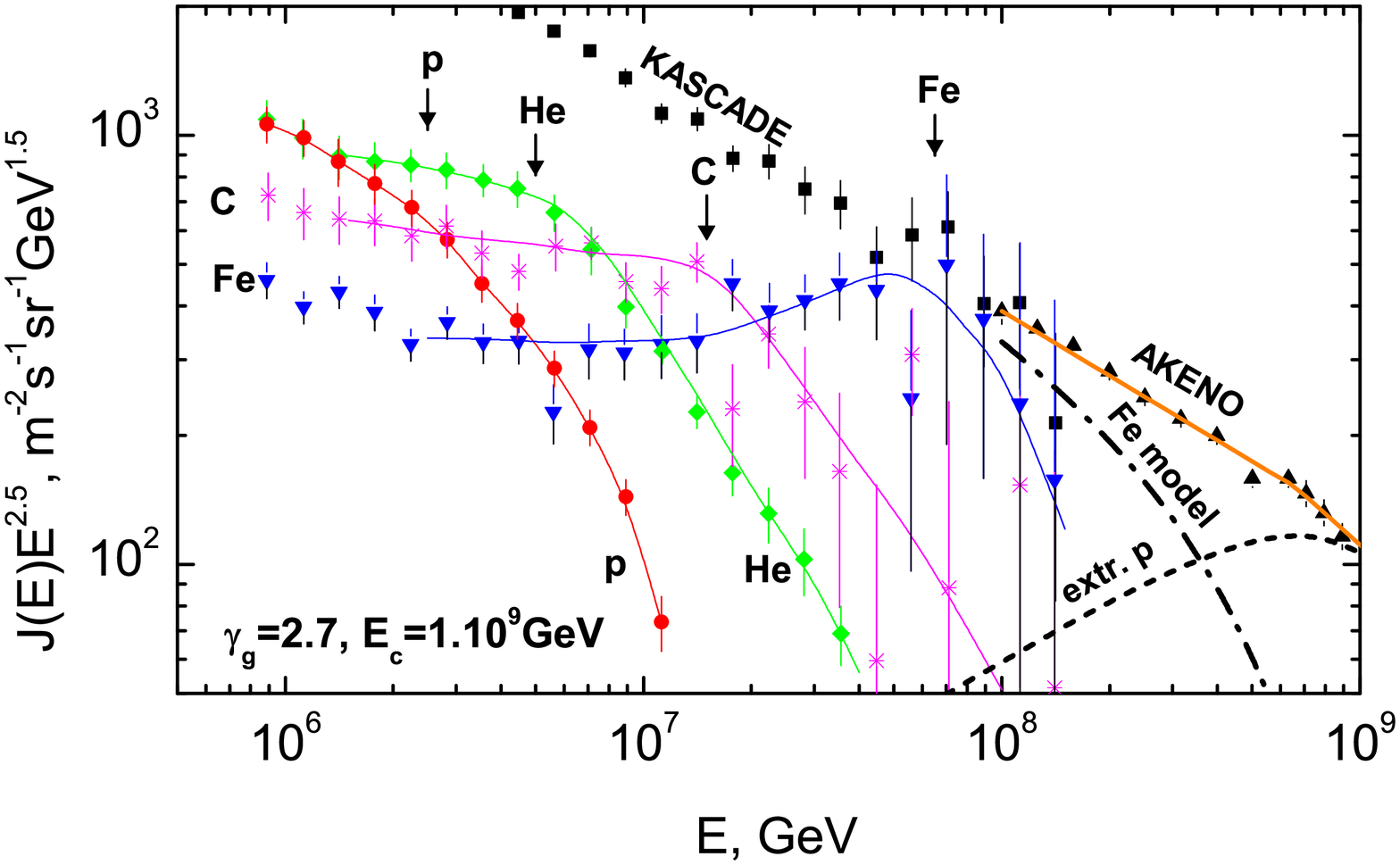}
\end{minipage}
\caption{\label{kascade} In the {\em left panel} the proton and nuclei
  spectra are shown according to KASCADE data. The arrows labelled by
  p, He, C and Fe show the positions of corresponding knees, calculated  
  as $E_Z=Z E_p$, with $E_p=2.5 \times 10^6$~GeV (the proton knee). One
  can notice the agreement between calculated and observed positions
  of the knees. In the {\em right panel} one can see the extragalactic
  proton spectrum, calculated in AGN model (see Section \ref{AGN}), and
  galactic iron-nuclei spectrum (Fe-model) predicted by this model. 
  }
\end{figure}

\subsection{Comparison of the dip and ankle transitions}
\label{fine-tuning}
{\em Dip transition} is based on rigidity model for galactic cosmic rays,
according to which the proton knee is at $E_p \approx 2.5 \times 10^{15}$~eV 
and iron knee is at $E_{\rm Fe} = ZE_p \approx 6.5\times 10^{16}$~eV. 
At $E > E_{\rm Fe}$ the galactic spectrum is predicted to have
steepening.  The physical spectra predicted in the rigidity model are
those for different nuclei, while the sum of these spectra has no
direct physical meaning. The total spectrum between $E_p$ and $E_{\rm Fe}$
must show some oscillations due to the knees of the different nuclei, 
and the power-law {\em total} spectrum 
$E^{-\gamma}$ at $E_p \leq E \leq E_{\rm Fe}$ is nothing more than a
fit, which cannot be extrapolated beyond $E_{\rm Fe}$. The question
whether or not continuation of $E^{-\gamma}$ spectrum at $E > E_{\rm Fe}$ 
fits the flux measured at $E > E_c=1\times 10^{18}$~eV is unphysical
within this model. 

The maximum acceleration energy $E_{\rm max}$ for galactic cosmic rays
may be in this model $E_{\rm max} < 1\times 10^{18}$~eV for iron
nuclei, which fits SN shock acceleration models.

At $E > E_{\rm Fe}$ the galactic iron flux steepens, while
extragalactic flux flattens at energies below $E_c=1\times 10^{18}$~eV  
(see curves 'Fe-model'' and ``extr. p' in the right panel of 
Fig \ref{kascade} ). These two curves thus inevitably intersect at
some energy $E_{\rm tr}$, providing the transition. The energy
behaviour of both curves and value of $E_{\rm tr}$ are model
dependent. 

In Fig \ref{kascade} (right panel) the transition is shown
for the model described in Section \ref{AGN}  The transition is
characterised by three energies 
$E_{\rm Fe}\approx 6.5\times 10^{16}$~eV,~ $E_{tr} \approx 4\times
10^{17}$~eV and $E_c \approx 1\times 10^{18}$~eV, and by fluxes 
$J_{\rm gal}(E_{\rm Fe}) \approx 1.8\times 10^{-17}~{\rm m}^{-2}{\rm s}^{-1}
{\rm sr}^{-1}{\rm GeV}^{-1}$ and 
$J_{\rm extr}(E_c) \approx 3.2\times 10^{-21}~{\rm m}^{-2}{\rm s}^{-1}
{\rm sr}^{-1}{\rm GeV}^{-1}$. It is difficult to find any fine-tuning
in these values. 
The flux at the 'end' of galactic spectrum, $E_{\rm Fe}$, and the
flux at the starting of extragalactic part of the spectrum, $E_c$, differ
by 4 orders of magnitude, and thus there is no conspiracy between
these two values.

The structure of the second knee, composed by galactic iron and
extragalactic protons (see right panel in Fig \ref{kascade}), is very
similar to the structures of the other knees: there is a sharp
transition from galactic iron to extragalactic protons (similar to 
transitions between different galactic nuclei), which in both cases
results in the total spectrum with a faint transition feature. 

The predicted spectrum shape above the second knee 
($1\times 10^{18} - 4\times 10^{19}$~eV) is model independent and is
confirmed by observations. The ankle at $E_a \sim 1\times 10^{19}$~eV
appears automatically in the calculations as the part of the dip.

The {\em ankle model} is based on the assumption that transition
occurs at ankle $E_a \sim 1\times 10^{19}$~eV, where experimental data
show flattening of the spectrum. An advantage of this model is given
by flatter generation spectrum ($\propto E^{-2.3}$), which allows to
diminish the source luminosities. The disadvantages are
connected with models for galactic cosmic rays at 
$E \lsim 1\times 10^{19}$~eV. What particles fill the gap between the iron
knee $E_{\rm Fe} \approx 6.5\times 10^{16}$~eV and the ankle 
$E_a \sim 1\times 10^{19}$~eV?  If they are iron nuclei, why they have 
the same spectrum as the {\em sum} of different nuclei at $E < E_{\rm Fe}$?
How protons, which start to disappear above the proton knee 
$E_p \approx 2.5\times 10^{15}$~eV re-appear again at energy 
$E > 1\times 10^{17}$~eV comprising about 10 \% of the Akeno flux? 
Why the spectrum at $1\times 10^{18}\leq E \leq 4\times 10^{19}$~eV
has the dip feature explained so well by extragalactic protons?

\section{The AGN model}
\label{AGN}
\begin{figure}[ht]
  \begin{center}
    \includegraphics[width=14.0cm]{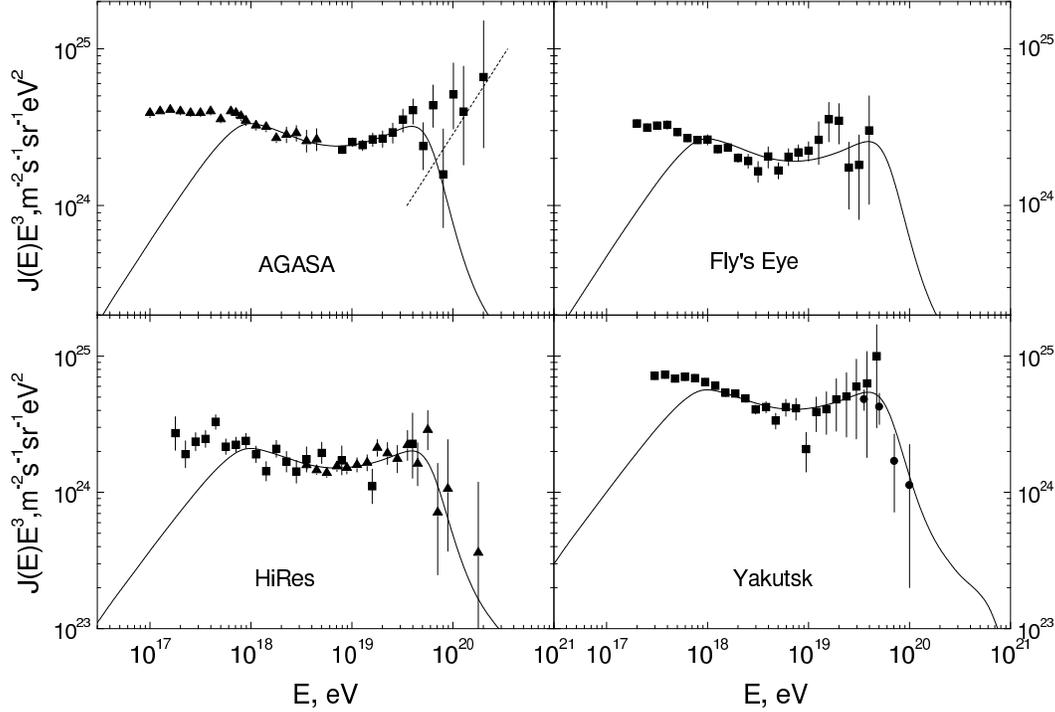}
\end{center}
 \caption{The AGN-model predicted spectra compared with data. 
}
  \label{spectra}
\end{figure}

The AGN as sources of UHECR meet the necessary requirements: 
{\em (i)} to provide the necessary energy output, {\em (ii)} to
have the space density $n_s \sim (1 - 3)\times 10^{-5}$~Mpc$^{-3}$ 
necessary for the observed small-scale anisotropy \cite{s-scale} 
and (iii)  to provide the observed correlations with BL Lacs \cite{TT}.
\\*[2mm]    
{\em Acceleration}\\*[2mm]
According to the AGN unified model BL Lacs are Fanaroff-Riley galaxies with 
jets in the direction of the observer. Thus, correlations of UHECR
particles with BL Lacs imply the jet acceleration of the particles. 
Acceleration in the jets due to pinch mechanism was suggested 
first for tokamaks (where it was confirmed on the laboratory scale)
and then was rescaled in \cite{pinch} to cosmic sizes. 
\begin{figure}[ht]
\begin{center}
\includegraphics[width=30mm,height=80mm,angle=89]{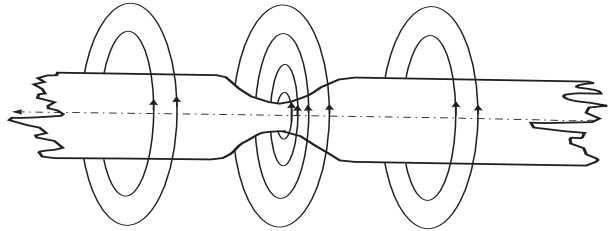}
\end{center}
\caption{Pinch-neck instability in jet}
\label{pinch}
\end{figure}
The pinch mechanism of acceleration works due to pinch-neck
instabilities, illustrated in Fig~\ref{pinch}, which are accompanied by
production of strong electric field. The solution of kinetic equations 
\cite{pinch} gives the power-law spectrum of accelerated particles 
$q(E) \propto E^{-\gamma_g}$~ with~ $\gamma_g=1 + \sqrt{3}$.
which is practically the same as in our calculations. The maximum
energy of acceleration can well exceed $10^{20}$~eV, if to  rescale
the laboratory pinch scale, where MeV energies were observed, to the
AGN size. \\*[2mm]
\noindent
{\em Generation spectrum and source luminosity}\\*[2mm]
Inspired by the pinch acceleration mechanism, we assume ad hoc the
generation spectrum in the form:
\begin{equation}
q_{\rm gen} (E) \propto \left\{ \begin{array}{ll} 
E_g^{-2}~~~&~~~~~~ {\rm at}~~ E_g \leq E_c\\
E_g^{-2.7}~&~~~~~~ {\rm at}~~ E_g \geq E_c ,\\
\end{array}
\right.
\label{gen-spectrum}
\end{equation}
with $E_c$ being a free parameter and 
with the maximum acceleration energy $E_{\rm max}=1\times 10^{21}$~eV.\\ 
For $E_c \sim 1\times 10^{18}$~eV the {\em emissivity} needed to fit
the calculations to the AGASA flux is 
${\cal L}_0=3.5\times 10^{46}$~erg/yr Mpc$^3$. With space density
of the sources taken from small-angle clustering 
$n_s \sim 3\times 10^{-5}$~Mpc$^{-3}$, the luminosity of a source is 
$L_p=3.7\times 10^{43}$~erg/s, which is quite low for AGN. \\*[2mm]
With these assumptions the calculated spectra in comparison with 
observations are shown in Fig~\ref{spectra}  The AGASA excess, if
real, needs for its explanation another component. shown by dashed line 
in the AGASA panel (e.g. from superheavy dark matter \cite{shdm}).

\vspace{2mm}
\noindent
{\em Transition from extragalactic to galactic cosmic rays}\\*[2mm]
Assuming that galactic cosmic rays at $E > 1\times 10^{17}$~eV are
dominated by iron nuclei, we calculate this flux subtracting the 
extragalactic proton flux, calculated in AGN model, from the measured
Akeno flux. The result is shown in Fig~\ref{kascade} (right panel). 


\section{Conclusions}
\label{conclusions}
\noindent
1. The {\em dip} is the spectrum feature produced by
interaction of UHE extragalactic protons with CMB radiation. In terms 
of modification factor it is
practically model-independent, depending weakly on generation index 
$\gamma_g$, mode of propagation (rectilinear or diffusive
propagation), discreteness and inhomogeneity in source distribution, 
local overdensity or deficit of the sources, maximum acceleration
energy and fluctuations in energy losses. For non-evolutionary sources,
when the number of free parameters is minimal ($\gamma_g$ and total
flux normalization) 
the agreement with observational data of Akeno-AGASA, HiRes, Fly's Eye 
and Yakutsk (see Fig~\ref{dips}) is very good and characterised by 
$\chi^2/d.o.f.$ equal to 1.12 for Akeno-AGASA and 0.908 for HiRes. 
It implies very low probability of accidental agreement.  
{\em This is the strong evidence that majority of primaries observed at 
$1\times 10^{18} - 4\times 10^{19}$~eV are extragalactic protons
propagating through CMB.}\\*[2mm]
2. The dip is modified by evolution of the sources and by presence of
extragalactic nuclei as the primaries. The latter effect
points to some acceleration mechanism or to the sources, where 
He-nuclei are photodisintegrated (see Section \ref{thedip}).\\*[2mm]
3. The dip has two flattenings, at low energy 
$E_c \approx 1\times 10^{18}$~eV and at high energy  
$E_a \approx 1\times 10^{19}$~eV. The latter, {\em ankle},  is 
confirmed in most observations (see Fig~\ref{dips}). The former
provides the dominance of more steep galactic component at $E < E_c$. 
There are three evidences of the transition from extragalactic to
galactic cosmic rays at energy $E_c$ shown in Fig~\ref{trans}  The
observed transition should be seen at energy $E < E_c$, and this feature
is observed as the {\em second knee}. The energy $E_c$,  where transition
from {\em galactic to extragalactic} cosmic rays is completed, is 
directly connected with energy $E_{\rm eq}$, where adiabatic and pair
production energy losses for extragalactic protons are equal.\\*[2mm] 
4. The transition from galactic to extragalactic cosmic rays at the {ankle}
is discussed in subsections \ref{ankle} and \ref{fine-tuning}  For the 
ankle transition one must 
assume that agreement of the dip with observations is accidental, while 
in case of the dip the ankle is an automatic feature (part of the dip).\\*[2mm]
5. The above-listed conclusions are valid for model-independent analysis
with basically two assumptions: the proton spectrum is power-law  at 
$E \gsim 1\times 10^{18}$~eV and sources are distributed uniformly 
with arbitrary separation $d$. The AGN model described in Section \ref{AGN}
uses some specific generation spectrum, inspired by pinch acceleration 
mechanism. This model explains small-scale anisotropy and correlations
with BL Lacs, predicts spectra of UHECR measured in Akeno-AGASA, HiRes,
Fly's Eye and Yakutsk experiments and quantitatively describes the
transition from extragalactic to galactic cosmic rays at the second
knee. The latter is quite similar to transitions from one type of
nuclei to another in the first knee. 

\section*{Acknowledgements}
I am grateful to Roberto Aloisio, Pasquale Blasi, Askhat Gazizov and 
Svetlana Grigorieva for most efficient and pleasant joint work, which 
results are (preliminary) presented here. 
We thank Transnational
Access to Research Infrastructure ILIAS-TARI through the contract 
HPRI-CT-2001-00149 within which most of work presented here has 
been performed. I am grateful to Vladimir Ptuskin, Oleg Smirnov and 
Alan Watson for valuable discussions.

\end{document}